\documentclass[12pt]{article}
\usepackage{graphicx}
\usepackage{cite}
\usepackage{epsfig,amsmath,amssymb} 
\usepackage{color}

\newcommand{\be}{\begin{equation}}
\newcommand{\ee}{\end{equation}}

\newcommand{\bit}{\begin{itemize}}
\newcommand{\eit}{\end{itemize}}
\newcommand{\bea}{\begin{eqnarray}}
\newcommand{\eea}{\end{eqnarray}}

\newcommand{\Neel}{N\'{e}el }

\usepackage{epsfig}

\begin{document}
\title
{The spin-$1/2$ $J_1$-$J_2$ Heisenberg antiferromagnet on the square lattice: 
Exact diagonalization for $N=40$ spins
}

\author
{J. Richter$^1$ and J.~Schulenburg$^2$
\\
\small{
$^{1}$Institut f\"ur Theoretische Physik, Universit\"at
Magdeburg, 39016 Magdeburg, Germany} \\
\small{
$^2$Universit\"{a}tsrechenzentrum,
             Universit\"{a}t Magdeburg, D-39016 Magdeburg, Germany}
}                     

\date{\today}
       
\maketitle

\begin{abstract}
We present numerical exact results for the ground state and the low-lying
excitations for the spin-$1/2$ $J_1$-$J_2$
Heisenberg antiferromagnet on finite square lattices of up to $N=40$ sites. 
Using finite-size extrapolation we determine the ground-state energy, the
magnetic order parameters, the spin gap, the uniform susceptibility, as well
as the spin-wave velocity and the spin stiffness as functions of the
frustration parameter $J_2/J_1$. In agreement with the generally excepted
scenario we find semiclassical magnetically ordered phases  for $J_2 < J_2^{c_1}$ and
$J_2 > J_2^{c_2}$ separated by a gapful quantum paramagnetic phase. We estimate
$J_2^{c_1} \approx 0.35J_1$ and 
$J_2^{c_2} \approx 0.66J_1$.    
\end{abstract}



75.45.+j Macroscopic quantum phenomena in magnetic systems

\section{Introduction}
The spin-$1/2$ Heisenberg
antiferromagnet (HAFM) with nearest-neighbor (NN) $J_1$ and frustrating
next-nearest-neighbor (NNN) $J_2$ coupling 
($J_1$-$J_2$ model) on the square lattice has attracted a
great deal of interest during the last twenty years (see, e.g., 
Refs.~\cite{chandra88,dagotto89,figu90,schulz,ivanov92,richter93,retzlaff,
richter94,Trumper97,voigt97,Einarsson95,
bishop98,singh99,sushkov01,capriotti00,capriotti01,siu01,singh03,roscilde04,
Sir:2006,Mam:2006,Schm:2006,Scheidl:2006,ueda07,Mune:2007,becca07,beach,bishop08,sousa,darradi08,ortiz,singh2009,ogata2009,cirac2009}
and references therein).
The corresponding Hamiltonian 
reads
\begin{equation}
H = J_{1}\sum_{\langle i,j
\rangle}{\bf s}_i \cdot {\bf s}_j
  +  J_{2}\sum_{\langle\langle i,k
\rangle\rangle}{\bf s}_i \cdot {\bf s}_j,
\end{equation}
where the sums over $\langle i,j \rangle$ and $\langle\langle i,k \rangle\rangle$ 
run over all NN and NNN pairs, respectively, counting each bond once. In what
follows we set $J_1=1$.
The synthesis of layered magnetic materials
\cite{melzi00,rosner02} which can be described by the spin-$1/2$ $J_1$-$J_2$ 
model has stimulated a renewed interest in this model.
Another new promising perspective is also opened by the recently discovered
layered Fe-based superconducting materials \cite{kamihara} which may have a
magnetic phase that can be described by a $J_1$-$J_2$ model with spin quantum
number $s>1/2$ \cite{yildirim,si,ma}.

The $J_1$-$J_2$ model is a canonical model to study quantum phase
transitions
between semiclassical magnetically ordered ground state (GS) phases and novel
non-magnetic quantum phases. Moreover, the  $J_1$-$J_2$ 
model might be also a candidate for a 
deconfined critical point separating a semiclassical magnetic phase from 
non-magnetic quantum phase
\cite{Sir:2006,darradi08,singh2009,Sent:2004a,Sent:2004b}.

Due to frustration highly efficient  quantum Monte-Carlo codes \cite{alps}, such as the
stochastic series expansion 
suffer from
the minus sign problem. 
Therefore, 
many other approximate methods, e.g., the Green's function
method \cite{siu01,Schm:2006}, the series
expansion \cite{singh99,singh03,sushkov01,Sir:2006,singh2009}, the Schwinger
boson approach \cite{Trumper97}, 
the coupled cluster method \cite{bishop98,Schm:2006,bishop08,darradi08}, 
variational techniques  \cite{capriotti01,beach}, the path integral
quantization \cite{Scheidl:2006}, the cluster effective-field
theory \cite{ueda07,sousa}, the hierarchical mean-field approach \cite{ortiz},
the projected entangled pair states method \cite{cirac2009}  as well as the  stochastic state
selection method \cite{Mune:2007}
were used to find the GS 
phases of the model.  
One of the most powerful methods is the Lanczos exact diagonalization
method
 which yields 
numerically exact results for finite lattices.
It has been succesfully applied to the $J_1$-$J_2$ model in various papers,
see, e.g., Refs.~\cite{dagotto89,figu90,schulz,richter93,retzlaff,richter94,
Einarsson95,voigt97,capriotti00,Mam:2006}.
In particular, the paper of Schulz and coworkers \cite{schulz} presenting for
the first time results for lattices up to $N=36$ sites has set a benchmark for
such studies and provides detailed information on the model. In addition,
these numerical exact results are often used to test new approximate
methods, see, e.g., Refs.~\cite{Kash:2001,Mune:2007,darradi08}.    

Due to the progress in the computer hardware and  also to the increase  
in 
the efficiency in programming the Lanczos algorithm, 
very recently  the GS and low-lying excitations
of the unfrustrated (i.e., $J_2=0$) spin-$1/2$ HAFM \cite{wir04} and of a spin-$1/2$
Heisenberg
model with ring exchange \cite{Lauch:2004} have been
calculated for a square lattice with $N=40$ sites.
The largest two-dimensional quantum spin model 
for which
the GS has been calculated so far is the spin-$1/2$ HAFM
on the star lattice with
$N=42$ sites \cite{star04}.
Note, however, that for higher sectors of $S_z$ which are relevant for finite magnetic
fields much larger system sizes can be considered, see, e.g., Ref.~\cite{prl02}.

\begin{figure}
\begin{center}
\includegraphics[clip=on,width=70mm,angle=0]{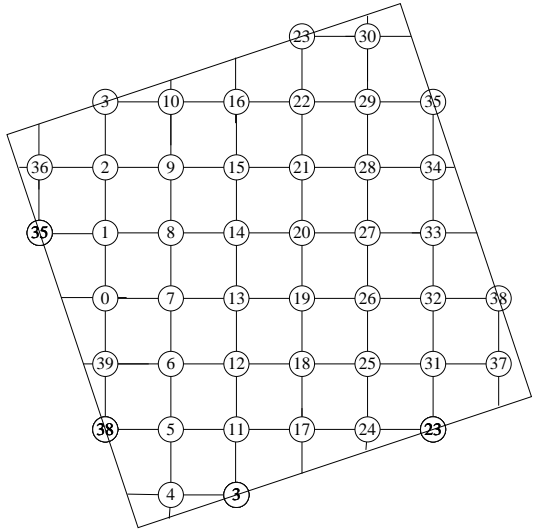}
\end{center}
\caption[]{The finite square lattice with $N=40$ sites.
\label{lattice}}
\end{figure}

As a result of numerous investigations it seems to be clear  now that
the $J_1$-$J_2$ model exhibits two magnetically long-range ordered phases 
at small and at  large $J_2$ 
separated by an intermediate  quantum paramagnetic phase 
without magnetic long-range order
(LRO)
in the parameter region 
$J_2^{c_1} < 
J_2 < J_2^{c_2}$, where $J_2^{c_1} \approx 0.4J_1$
and  $ J_2^{c_2} \approx 0.6J_1$.
The GS at low $J_2 < J_2^{c_1}$ 
exhibits semiclassical N\'eel magnetic LRO with the
magnetic wave vector  
  ${\bf Q}_{0}=(\pi ,\pi )$.
The GS at large $J_2 > J_2^{c_2} $ shows so-called 
collinear magnetic LRO with  the 
magnetic wave vectors ${\bf Q}_{1} =(\pi , 0)$
or ${\bf Q}_{2} =(0 ,\pi )$. 
These two collinear states are characterized by a parallel spin orientation of 
nearest neighbors in vertical (horizontal) direction and an antiparallel 
spin orientation of  nearest neighbors in horizontal (vertical) 
direction.

The nature of the transition
between the N\'eel and the quantum paramagnetic phase 
as well as the properties of the quantum paramagnetic phase and the 
precise values of the transition points are  
still under
discussion \cite{chandra88,dagotto89,figu90,schulz,ivanov92,richter93,retzlaff,
richter94,Einarsson95,bishop98,
singh99,sushkov01,capriotti00,capriotti01,siu01,singh03,roscilde04,
Sir:2006,Mam:2006,Schm:2006,Scheidl:2006,ueda07,Mune:2007,becca07,beach,bishop08,sousa,darradi08,ortiz,singh2009}.

In this paper we present new results for the GS and the low-lying
excitations of the
$J_1$-$J_2$ model
for a finite lattice of $N=40$ sites shown as in Fig.~\ref{lattice}. These results
are presented in detail, on the one hand, as benchmark data for approximate
methods. On the other hand, we combine our new results with the known
results \cite{schulz}
(which we have recalculated) to improve the finite-size extrapolation. For
all the finite lattices periodic boundary conditions were imposed.

\section{Results of Lanczos exact diagonalization for a finite square lattice
of $N=40$ sites} 
Although Schulz and coworkers performed the Lanczos diagonalization for the
spin-$1/2$ $J_1$-$J_2$ HAFM 
on the finite square lattice of $N=36$ sites more then 10 years ago the corresponding Lanczos
diagonalization for
the finite lattice of $N=40$ sites is still a challenging problem.
The number of basis states 
on the square lattice in the GS symmetry sector is $n_h=430 909
650$ for $N=40$ compared to $n_h=15804956$ for $N=36$.
The GS belongs to a translational symmetry with ${\bf q}=(0,0)$.   

The most important quantities in order to analyze semiclassical GS 
magnetic ordering are appropriate  
order parameters corresponding to the classical \Neel and collinear LRO.
Following Schulz et al. \cite{schulz} we use here 
the ${\bf Q}$-dependent susceptibilities
(square of order parameters) defined as
 \be  \label{M_squared}
M_N^2({\bf Q})= 
\frac{1}{N(N+2)}\sum_{i,j} \langle{\bf s}_i \cdot {\bf s}_j\rangle 
e^{i{\bf Q}({\bf R}_i - {\bf R}_j)}.
\ee
For the magnetic wave vector  
  ${\bf Q}_{0}=(\pi ,\pi )$ the quantity $M_N^2({\bf Q_0})$ is the 
relevant order parameter for the \Neel ordered GS phase present at small
$J_2$, whereas $M_N^2({\bf Q_{1(2)}})$ 
with magnetic wave vectors ${\bf Q}_{1} =(\pi , 0)$
or ${\bf Q}_{2} =(0 ,\pi )$
corresponds to the collinear magnetic LRO present at large $J_2$.

In table \ref{tab1} we give the singlet GS energy, $E_{GS}(S=0)$, the energy
of the first triplet excitation $E_0(S=1)$, the energy of the first singlet 
excitation $E_1(S=0)$, as well as the ${\bf Q}$-dependent susceptibilities
$M_N^2(\pi,\pi)$,
$M_N^2(\pi,0)$.
The low-lying energies are also displayed in
Fig.~\ref{ener40} where more data points are included than shown in
table~\ref{tab1}. 
It is obvious that the first excited singlet state becomes very close to the
singlet GS near     
$J_2=0.6$. Since both states
belong to the same lattice symmetry, a level crossing is forbidden due to the
von Neumann-Wigner theorem \cite{neumann}. The numerical data shown in the right inset
of Fig.~\ref{ener40} clearly give a numerical check of this general theorem.
The energies of the low-lying states show the typical frustration induced maximum
known from previous ED calculations for smaller lattices, see, e.g., 
Refs.~\cite{dagotto89,schulz,richter93}. 
Since we have data for the singlet GS and the first triplet excitation, we
can also give results for the spin gap 
$\Delta_T=E_0(S=1)-E_{GS}(S=0)$, see Fig.~\ref{gap}.
Note that the spin gap was not presented in the paper of Schulz et
al.\cite{schulz}. Hence we compare in Fig.~\ref{gap} spin gap data for $N=32$, $36$,
and  $40$. 
It is obvious that with increasing $J_2$ the gap starts to grow at
$J_2 \approx 0.4 J_1$. It reaches a maximum at about $J_2 \approx 0.58 J_1$ and becomes
again small for $J_2 \gtrsim 0.6 J_1$.
This behavior yields an indication of a gapful quantum paramagnetic GS phase
around $J_2=0.5J_1$.
However, it is
also obvious that there is no monotonous  finite-size behavior  in that
parameter region, i.e., a
reliable finite-size extrapolation of the spin gap around $J_2=0.5J_1$ is not possible,

Next we consider the spin-spin correlation functions $\langle{\bf s}_0 \cdot {\bf
s}_{\bf R}\rangle$. There are altogether 11 different correlations functions given
in table~{\ref{tab2} for the same data points as in table~\ref{tab1}.      
Graphically the variation of some selected correlation functions are shown
inf Fig.~\ref{sisj40}. For small $J_2$ the spin-spin correlations
are quite strong and according to the \Neel order we have  $\langle{\bf s}_0 \cdot {\bf
s}_{\bf R}\rangle <0 $ for $\bf R$ connecting sites of sublattices $A$ and $B$
but $\langle{\bf s}_0 \cdot {\bf
s}_{\bf R}\rangle >0 $ for $\bf R$ connecting sites within a sublattice.
Increasing the  frustration leads to a weakening of the spin-spin
correlation. This weakening is particularly strong for larger separations. 
For strong frustration around
$J_2=0.5J_1$ all correlation functions $\langle {\bf s}_0 \cdot {\bf s}_{\bf
R}\rangle$ except for ${\bf R}=(1,0);\;(1,1);\;(2,0)$ are very small
which is an indication for a magnetically disordered phase.  
Beyond $J_2 \approx 0.6J_1$ the increasing strength of correlation functions
for lattice vectors $\bf R$ connecting
sites within the same sublattice $A$ (or $B$) indicates the emerging
collinear LRO.
The corresponding order parameters defined in Eq.~(\ref{M_squared}) and listed
in table~\ref{tab1} are shown in Fig.~\ref{mag40}. The suppression of magnetic
order around $J_2=0.5J_1$ is again obvious.
Although there is no level crossing (see the discussion above) 
we observe in Figs.~\ref{sisj40} and \ref{mag40} a sharp
change of the magnetic quantities near $J_2=0.6J_1$, i.e., at that point where
both singlet levels become very close to each other (see Fig.~\ref{ener40}, right inset). 
This behavior might be
a hint on a first-order transition between the phase with collinear LRO and
the magnetically disordered phase, see also, e.g., 
Refs.~\cite{schulz,singh99,sushkov01,Schm:2006,darradi08}.

\begin{table}
\begin{center}
\caption
{Ground state energy $E_{GS}(S=0)$, first triplet excitation energy $E_0(S=1)$, first
singlet excitation energy  $E_1(S=0)$, square of \Neel order parameter  $M_N^2(\pi,\pi)$
and square of collinear order parameter $M_N^2(\pi,0)$. 
\label{tab1}}
\vspace{5mm}
\begin{tabular}[t]{|c||c|c|c|c|c|c|c|c|c|c|c|c|c|c|c|} \hline
 $J_2$ & $E_{GS}(S=0)$ & $E_0(S=1)$ & $E_1(S=0)$ & $M_N^2(\pi,\pi)$ & $M_N^2(\pi,0)$ \\ \hline \hline
 0.0 &  -27.09485025 & -26.83322962  & -26.3138435   & 0.193923  & 0.011956 \\
 0.10&  -25.46460260 & -25.19683764  & -24.6669636   & 0.184079  & 0.011994 \\   
 0.20&  -23.90046918 & -23.62413648  & -23.0809149   & 0.171482  & 0.012066 \\ 
 0.30&  -22.42728643 & -22.13740165  & -21.5763396   & 0.154648  & 0.012212 \\ 
 0.40&  -21.08836670 & -20.77332213  & -20.1877770   & 0.130935  & 0.012547 \\ 
 0.50&  -19.96304839 & -19.59762345  & -19.0413364   & 0.097703  & 0.013446 \\ 
 0.55&  -19.51791526 & -19.11727950  & -18.8522256   & 0.079043  & 0.014457 \\
 0.60&  -19.18368038 & -18.90110215  & -19.1762893   & 0.020304  & 0.078160 \\
 0.65&  -20.04603255 & -19.75459787  & -19.3898477   & 0.013516  & 0.093232 \\
 0.70&  -21.05530239 & -20.75925837  & -20.2366849   & 0.010120  & 0.101552 \\
 0.80&  -23.34020427 & -23.01691919  & -22.5277370   & 0.005522  & 0.111045 \\
 0.90&  -25.83691287 & -25.46196339  & -24.8738325   & 0.002912  & 0.115628 \\
 1.00&  -28.43880892 & -28.00735838  & -27.2299736   & 0.001700  & 0.117879 \\
\hline
\end{tabular}                                            
\end{center}
\end{table}

\begin{table}
\begin{center}
\caption
{Spin-spin correlation function $\langle {\bf s}_0 \cdot {\bf s}_{\bf R}\rangle$
(respectively, $\langle {\bf s}_0 \cdot {\bf s}_{i}\rangle$, see Fig.~\ref{lattice})
\label{tab2}}
\vspace{5mm}
\begin{tabular}[t]{|c||c|c|c|c|c|c|c|c|c|c|c|c|c|c|c|} \hline
       & $\langle {\bf s}_0\cdot{\bf s}_1\rangle$ & $\langle {\bf s}_0\cdot{\bf s}_2\rangle$ & $\langle {\bf s}_0\cdot{\bf s}_3\rangle$ & $\langle {\bf s}_0\cdot{\bf s}_4\rangle$ & $\langle {\bf s}_0\cdot{\bf s}_5\rangle$ & $\langle {\bf s}_0\cdot{\bf s}_6\rangle$  \\ \hline \hline
 $J_2$ & ${\bf R}=(1,0)$ & ${\bf R}=(2,0)$ & ${\bf R}=(3,0)$ & ${\bf R}=(3,1)$ & ${\bf R}=(2,1)$ & ${\bf R}=(1,1)$  \\ \hline \hline
 0.0   & -0.338685 &   0.185133 &  -0.166989 &   0.164100 &  -0.179037 &   0.207198  \\ 
 0.10  & -0.338313 &   0.175458 &  -0.154968 &   0.151833 &  -0.166128 &   0.200046  \\    
 0.20  & -0.336852 &   0.163089 &  -0.140055 &   0.136281 &  -0.149688 &   0.190488  \\  
 0.30  & -0.333423 &   0.146655 &  -0.120879 &   0.115776 &  -0.127803 &   0.176943  \\  
 0.40  & -0.326073 &   0.123909 &  -0.095208 &   0.087441 &  -0.097056 &   0.156171  \\  
 0.50  & -0.310788 &   0.093354 &  -0.061983 &   0.048666 &  -0.053772 &   0.122499  \\  
 0.55  & -0.298596 &   0.077397 &  -0.045300 &   0.027102 &  -0.028947 &   0.099315  \\ 
 0.60  & -0.129468 &   0.145560 &  -0.032355 &  -0.127482 &   0.034386 &  -0.183879  \\ 
 0.65  & -0.096630 &   0.168312 &  -0.027747 &  -0.156408 &   0.033291 &  -0.236841  \\ 
 0.70  & -0.077052 &   0.181365 &  -0.024825 &  -0.170640 &   0.030228 &  -0.265914  \\ 
 0.80  & -0.050388 &   0.195918 &  -0.020247 &  -0.186357 &   0.025323 &  -0.301704  \\ 
 0.90  & -0.034740 &   0.202785 &  -0.016548 &  -0.193845 &   0.021465 &  -0.320247  \\ 
 1.00  & -0.026295 &   0.206295 &  -0.013959 &  -0.197463 &   0.018474 &  -0.329190  \\
 \hline 
\end{tabular}

\vspace{0.1cm}
\begin{tabular}[t]{|c||c|c|c|c|c|c|c|c|c|c|c|c|c|c|c|} \hline
       & $\langle {\bf s}_0\cdot{\bf s}_9\rangle$ & $\langle {\bf s}_0\cdot{\bf s}_{10}\rangle$ & $\langle {\bf s}_0\cdot{\bf s}_{11}\rangle$ & $\langle {\bf s}_0\cdot{\bf s}_{16}\rangle$ & $\langle {\bf s}_0\cdot{\bf s}_{24}\rangle$  \\ \hline \hline
 $J_2$ & ${\bf R}= (1,2)$  & ${\bf R}= (1,3)$ & ${\bf R}= (2,2)$ & ${\bf R}= (2,3)$ & ${\bf R}= (4,-2)$  \\ \hline \hline
 0.0   & -0.177165 &  0.157032 &   0.161478 &  -0.156222 &  0.150825 \\
 0.10  & -0.164196 &  0.144276 &   0.148815 &  -0.142812 &  0.137613 \\   
 0.20  & -0.147774 &  0.128166 &   0.132678 &  -0.125913 &  0.120888 \\ 
 0.30  & -0.126123 &  0.107007 &   0.111225 &  -0.103674 &  0.098736 \\ 
 0.40  & -0.096186 &  0.077940 &   0.081246 &  -0.072885 &  0.067683 \\ 
 0.50  & -0.055281 &  0.039045 &   0.040050 &  -0.031119 &  0.024636 \\ 
 0.55  & -0.032856 &  0.018723 &   0.018048 &  -0.009279 &  0.001776 \\
 0.60  &  0.015885 & -0.109425 &   0.107499 &   0.004956 &  0.092322 \\
 0.65  &  0.017844 & -0.138945 &   0.137898 &   0.002280 &  0.124962 \\
 0.70  &  0.017595 & -0.155349 &   0.155199 &   0.000924 &  0.142587 \\
 0.80  &  0.016110 & -0.173610 &   0.174066 &   0.000216 &  0.159999 \\
 0.90  &  0.014244 & -0.182238 &   0.182661 &   0.000291 &  0.166980 \\
 1.00  &  0.012522 & -0.186480 &   0.186960 &   0.000333 &  0.170301 \\
\hline
\end{tabular}
\end{center}
\end{table}

\begin{figure}
\begin{center}
\includegraphics[clip=on,width=88mm,angle=270]{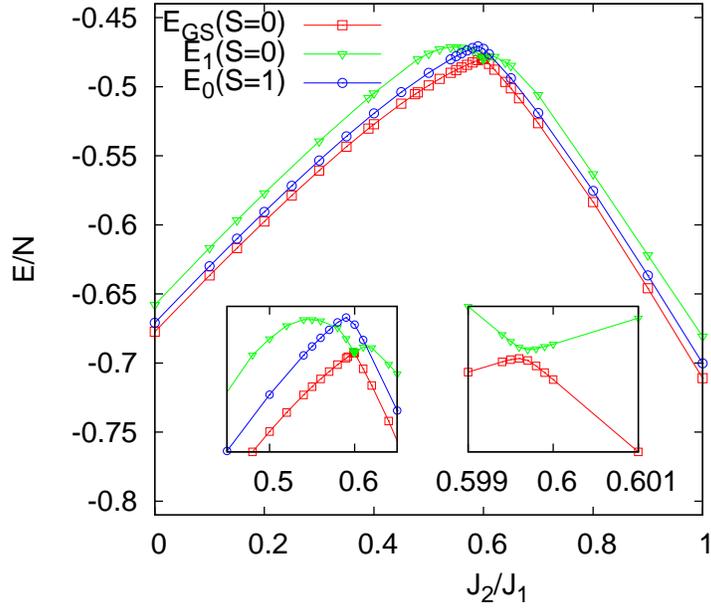}
\end{center}
\caption{Singlet GS energy, $E_{GS}(S=0)$, the energy
of the first triplet excitation, $E_0(S=1)$, and the 
energy of the first singlet
excitation, $E_1(S=0)$, versus $J_2$ for the square lattice with $N=40$ sites.
}
\label{ener40}
\end{figure}
\begin{figure}
\begin{center}
\includegraphics[clip=on,width=88mm,angle=270]{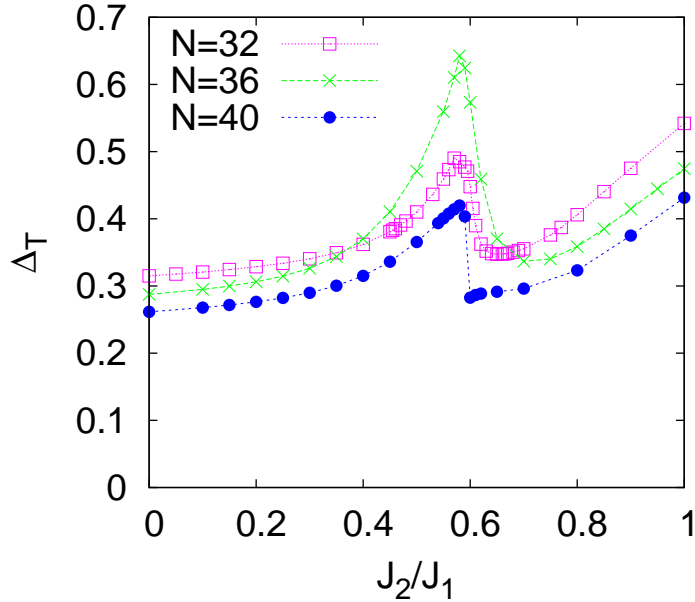}
\end{center}
\caption{Spin gap $\Delta_T= E_{GS}(S=1)-E_0(S=0)$ versus $J_2$ for $N=32$, $36$, and $40$.}
\label{gap}
\end{figure}

\begin{figure}
\begin{center}
\epsfig{file=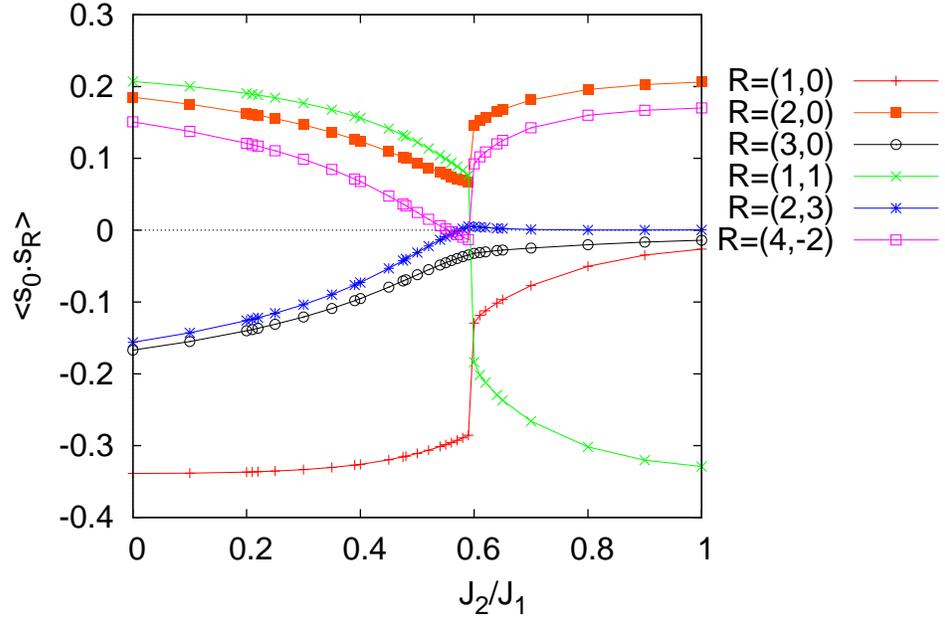,scale=0.5,angle=270.0}
\end{center}
\caption{ Selected spin-spin correlation functions  $\langle {\bf s}_0
\cdot {\bf s}_{\bf R}\rangle $ for $N=40$.
}
\label{sisj40}
\end{figure}

\begin{figure}
\begin{center}
\epsfig{file=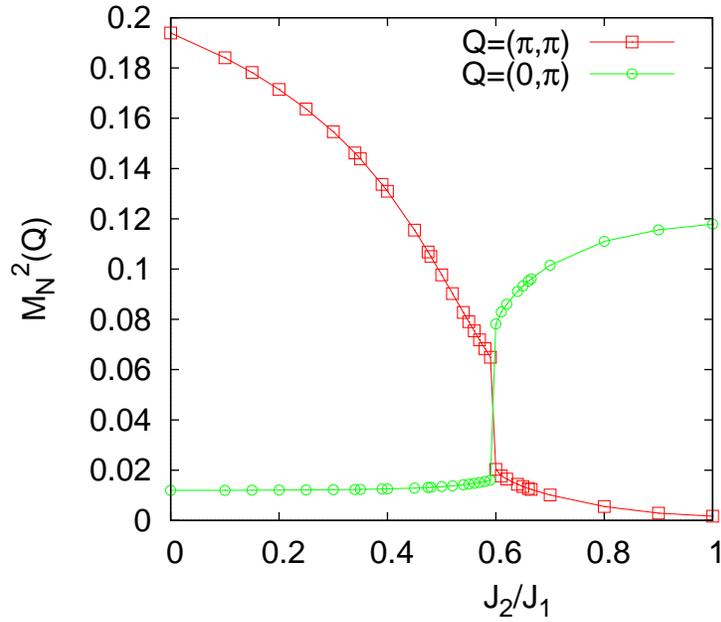,scale=0.5,angle=270.0}
\end{center}
\caption{ Square of order parameters $M_N^2({\bf Q})$ for ${\bf
Q}=(\pi,\pi)$ and ${\bf
Q}=(0,\pi)$, see Eq.~(\ref{M_squared}).}
\label{mag40}
\end{figure}
\begin{figure}
\begin{center}
\epsfig{file=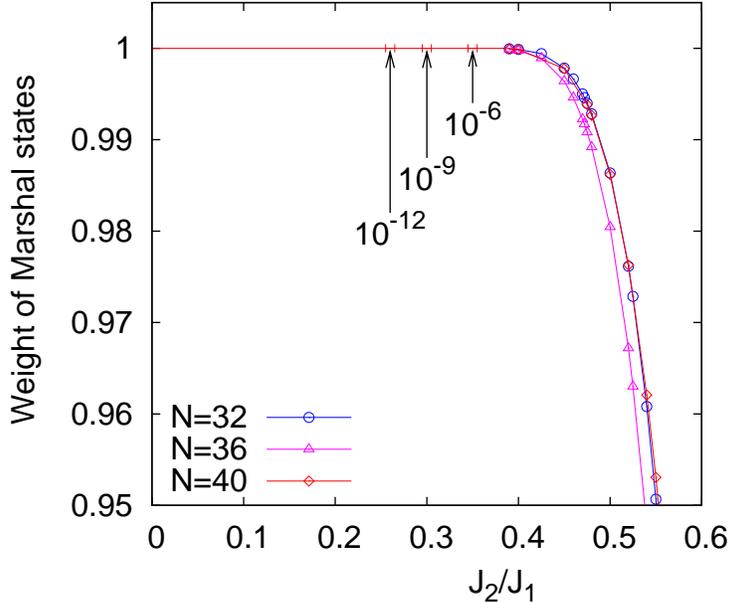,scale=0.5,angle=270.0}
\end{center}
\caption{Validity of the MPSR:
The curve shows the weight of the states fulfilling the MPSR, i.e., $\sum'_n
|c_n|^2$ (where the sum $\sum'$ runs over those states only, for which the
sign of $c_n$ fulfills the MPSR). The arrows indicate
those $J_2$ values for $N=40$, where the weight of the states which do not fulfill the
MPSR is $10^{-12}$, $10^{-9}$ and $10^{-6}$,
respectively.
}
\label{mrule}
\end{figure}

Another interesting point is the breakdown of the Marshall-Peierls sign rule
(MPSR)
at $J_2 \approx 0.4J_1$. 
Writing the GS as $|\Psi\rangle=\sum_n c_n |n\rangle$, where $|n\rangle$ is
an Ising basis state of typical form
$|\uparrow\uparrow\downarrow\uparrow\downarrow\downarrow\cdots\rangle$, the MPSR 
determines the sign of the coefficients $c_n$~\cite{marshall55}. The MPSR has been
proved exactly
for bipartite lattices and arbitrary site spins by Lieb, Schultz and
Mattis~\cite{lieb61}. As pointed out in several papers the knowledge of
the sign of the $c_n$ is of great importance in different numerical methods, e.g.
for the construction of variational wave functions
\cite{retzlaff,capriotti01,gros,beach}, in quantum Monte-Carlo methods (which suffer from
the sign problem in frustrated systems
\cite{readt81}) and also in the density matrix renormalization group
method, where the application of the MPSR has substantially improved
the method in a frustrated spin system \cite{schollwoeck98}.

For the
$J_1$-$J_2$ model on the square lattice the violation of the MPSR was
considered as one signal of destruction of magnetic LRO
\cite{retzlaff,richter94,voigt97,bishop98,capriotti01,beach}. 
It was found 
that the
MPSR (proved exactly
only for $J_2=0$)
survives till about
$J_2 \approx 0.3 \ldots 0.4J_1$.
The exact Lanczos results for the ground state 
presented in Refs.~\cite{retzlaff,richter94} and 
\cite{voigt97}
were, however, restricted to systems of up to $N=24$ sites.
Here we present data for $N=32$, $36$, and $40$ in Fig.~\ref{mrule}. 
From Fig.~\ref{mrule} it is obvious that the MPSR is
valid in very good approximation even till $J_2\approx 0.4J_1$. The MPSR starts
to be significantly violated beyond $J_2 = 0.4J_1$, where the \Neel LRO breaks down.

\begin{figure}
\begin{center}
\epsfig{file=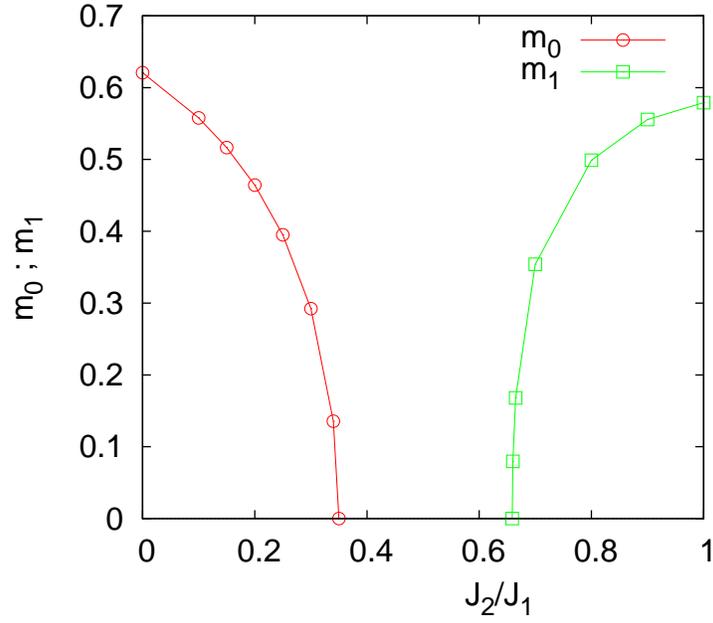,scale=0.5,angle=270.0}
\end{center}
\caption{Order parameters $m_0$ (see Eq.~(\ref{m_0})) and $m_1$ (see
Eq.~(\ref{m_1})) as a function of $J_2/J_1$.}
\label{extra_order}
\end{figure}

\begin{figure}
\begin{center}
\epsfig{file=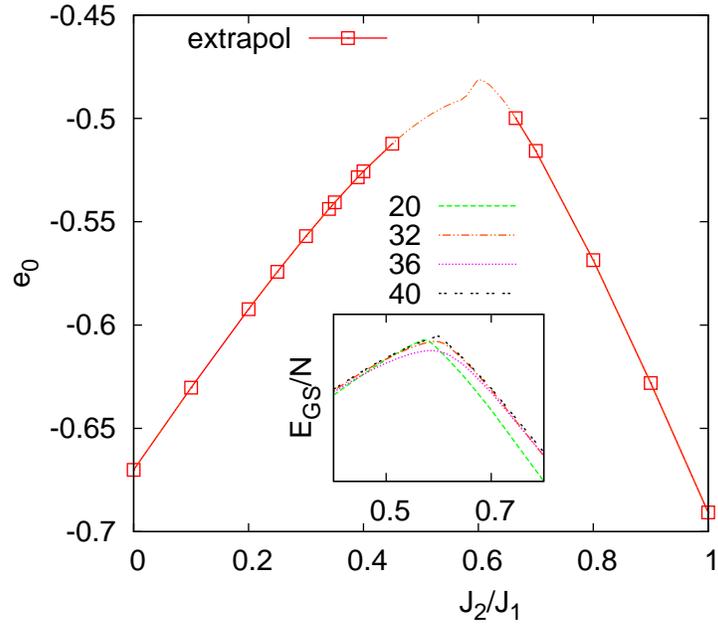,scale=0.5,angle=270.0}
\end{center}
\caption{Ground-state energy per site. Main panel: Extrapolated value
$e_0$
(see Eq.~(\ref{e}). Inset: Data for $N=20$, $32$, $36$, and $N=40$ around
the maximum. }
\label{E_extra}
\end{figure}

\section{Finite-size extrapolation}
We use now our new data for $N=40$ together with the known data for $N=20,
32, 36$ to perform a finite-size extrapolation. Note that we do not include the data
for the lattice with $N=16$ sites.
This lattice  has an extra symmetry because it is equivalent to a hypercube
in 4 dimensions. It was argued  that, therefore,  the $N=16$ lattice exhibits
an anomalous behavior \cite{schulz}. The finite-size extrapolation rules for 
the two-dimensional HAFM are well known
\cite{neuberger,hasenfratz,sandvik,betts1,betts1a}. 
The extrapolation of the magnetic order parameters 
yields an estimate of the transition points
between the semiclassically ordered phases (\Neel and collinear) and the
magnetically disordered quantum phase.
In addition, since the scaling behavior is related to
the low-energy degrees of freedom of the model,  via the
extrapolation procedure one can  extract the $J_2$ dependence of other quantities 
such as the spin-wave velocity $c$
and the spin stiffness $\rho$.
It has been demonstrated that Lanczos data for finite lattices of up to N=36 sites
can provide quite accurate data for these quantities in the case of
unfrustrated lattices \cite{schulz,betts1,betts2}.

In the seminal paper 
of Schulz et al. \cite{schulz} the finite-size
extrapolation procedure as well as some particular problems of the extrapolation appearing for
the $J_1$-$J_2$ model were discussed in great detail. We follow here the lines of that paper, but
do not repeat a detailed discussion of the extrapolation scheme.
   
We define the staggered magnetization, i.e., the \Neel order parameter, as
$m_0 = 2 \lim_{N \to \infty} M_N(\pi,\pi)$.   
The finite-size behavior of  $M_N(\pi,\pi)$ is given
by \cite{schulz,neuberger,hasenfratz,sandvik}
\be \label{m_0} 
M_N^2(\pi,\pi)=
 \frac{1}{4}m_0^2\left(
1 + \frac{0.62075 c}{\rho \sqrt{N}} + \cdots \right )
\ee
where $c$ is the spin-wave velocity and $\rho$ is the spin stiffness.
The corresponding order parameter for the collinear LRO at large $J_2$ is 
defined as \cite{schulz}
$m_1 = \sqrt{8} \lim_{N \to \infty} M_N(\pi,0)$.   
The finite-size behavior of  $M_N(\pi,0)$ is 
\be \label{m_1} 
M_N^2(\pi,0)= \frac{1}{8}m_1^2 + \frac{\rm const.}{\sqrt{N}} + \cdots \; .
\ee
The results for $m_0$ and $m_1$ are shown in Fig.~\ref{extra_order}. As
expected we find a magnetically disordered GS phase around $J_2= 0.5J_1$.   
The transition points are determined to $J_2^{c_1}=0.35J_1$ and
$J_2^{c_2}=0.66J_1$, i.e., the range of the magnetically disordered GS phase
obtained here is 
slightly larger than predicted, e.g., from series
expansion \cite{singh99,sushkov01,singh03,Sir:2006} or coupled
cluster approach \cite{darradi08}.
For the unfrustrated square lattice HAFM ($J_2=0$) we obtain $m_0=0.621$
which is in good agreement with corresponding results obtained by other
methods,
e.g., 3rd order spin-wave theory  ($m_0=0.6138$) \cite{hamer}, quantum
Monte Carlo ($m_0=0.6140$) \cite{qmc} or coupled cluster method
($m_0=0.6205$) \cite{ccm}.

The finite-size behavior of  the GS energy is given by 
\be \label{e} 
\frac{E_{GS}(N)}{N}= e_0 - 1.4372\frac{c}{N^{3/2}} + \cdots \; .
\ee
The extrapolated energy $e_0$ is shown in Fig.~\ref{E_extra}. Because of an
irregular finite-size behavior near the
maximum in $e_0$ (see inset in Fig.~\ref{E_extra}) the extrapolation becomes  unreliable
around $J_2=0.5J_1$ (this parameter region is indicated by the dotted line in the main panel of
Fig.~\ref{E_extra}).  Nevertheless, one can speculate that the kink in the
extrapolated GS energy near $J_2=0.6J_1$ might be another hint for a first-order
transition at $J_2^{c_2}$ between the semiclassical collinear phase and the
quantum paramagnetic phase.
The extrapolated GS energy, $e_0=-0.6701$, for $J_2=0$ 
is again in very good agreement with corresponding results obtained by other
methods, e.g., 3rd order spin-wave theory ($e_0=-0.66931$) \cite{hamer}, quantum
Monte Carlo ($e_0=-0.66944$) \cite{qmc} or coupled cluster method
($e_0=-0.66936$) \cite{ccm}.

\begin{figure}
\begin{center}
\epsfig{file=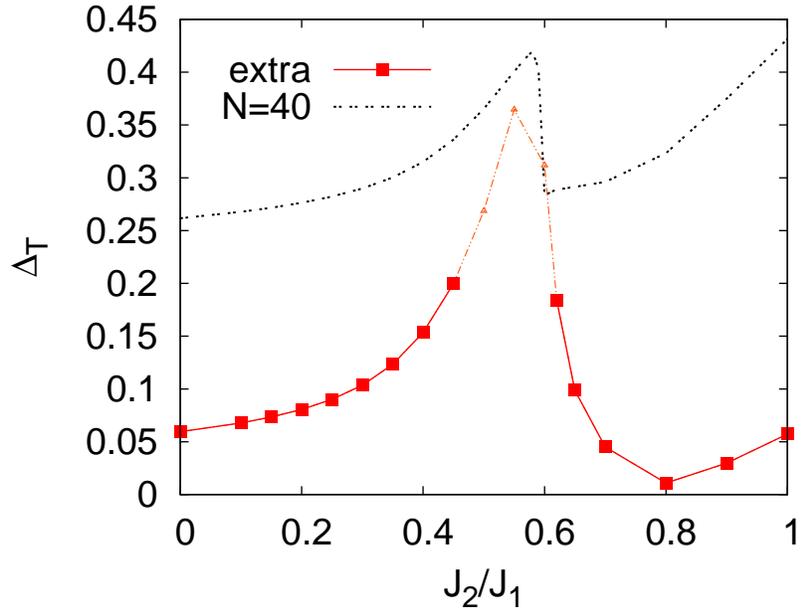,scale=0.5,angle=270.0}
\end{center}
\caption{Spin gap $\Delta_T$ for $N=40$ and  extrapolated to $N\to \infty$
(see Eq.~(\ref{eq_gap})). 
}
\label{gap_extra}
\end{figure}

\begin{figure}
\begin{center}
\epsfig{file=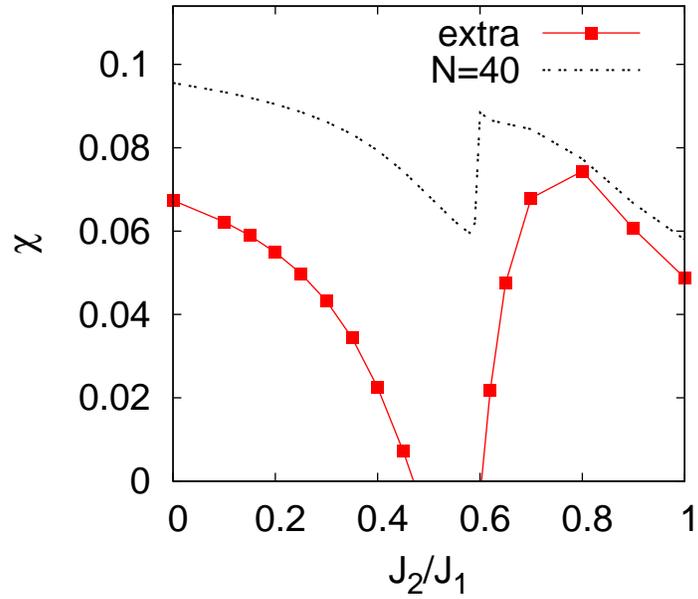,scale=0.5,angle=270.0}
\end{center}
\caption{Susceptibility $\chi$ determined according to Eq.~(\ref{eq_chi}).}
\label{chi_extra}
\end{figure}

\begin{figure}
\begin{center}
\epsfig{file=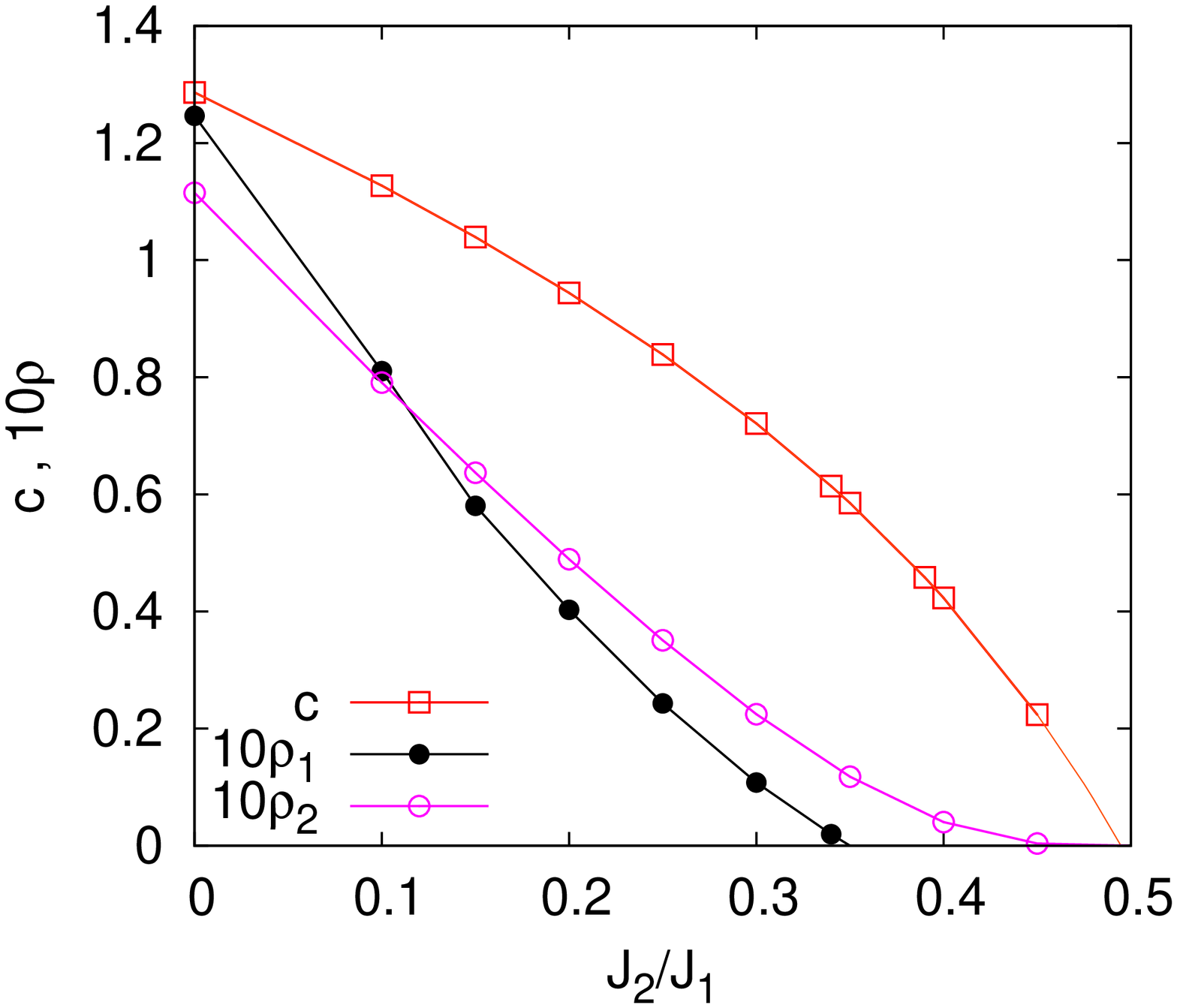,scale=0.5,angle=270.0}
\end{center}
\caption{Spin wave velocity $c$  and spin stiffness
$\rho$. For better comparison $\rho$ is multiplied by $10$. }
\label{c_sw}
\end{figure}

We can also perform a finite-size extrapolation of the gap 
to the first triplet excitation (spin
gap) $\Delta_T(N)=E_0(S=1)-E_{GS}(S=0)$. The corresponding formula is 
\be \label{eq_gap} 
\Delta_T(N) = \Delta_T + \frac{a}{N} 
+ \cdots \; .
\ee
The results for $\Delta_T$ are shown in
Fig.~\ref{gap_extra}. In the magnetically ordered  \Neel and collinear
phases gapless Goldstone modes exist and consequently the spin gap should
vanish. The finite values of the extrapolated gap for $J_2 < J_2^{c_1}$ and
$J_2 > J_2^{c_2}$ give a hint of the limits of precision of the finite-size
extrapolation of the spin gap.  Nevertheless, it becomes obvious   
that there is a significant increase of $\Delta_T$ in the quantum
paramagnetic phase   $J_2^{c_1} < J_2 <
J_2^{c_2}$. Note that for larger $J_2$ the energy scale of the system 
becomes proportional to $J_2$ that explains the  increase in $\Delta_T$ with $J_2$ for
$J_2>0.8$. A finite spin gap for $J_2^{c_1} < J_2 <
J_2^{c_2}$  would be consistent with the findings of many papers which
suggest that the quantum
paramagnetic phase exhibits valence-bond order, see, e.g.,
Refs.~\cite{Sir:2006} and \cite{darradi08} and references therein. Note, however,
that our finding is in contrast to recent DMRG studies on frustrated odd-leg
ladders \cite{becca07}
where arguments for a gapless spectrum of the $J_1$-$J_2$ model on the square
lattice have been given.

Next we consider the uniform susceptibility $\chi$ which is related to the
spin gap by $\chi(N)=N \Delta_T(N)$. 
For the finite-size extrapolation of the uniform susceptibility we
use \cite{schulz}
\be \label{eq_chi} 
\chi(N) = \chi + \frac{b}{\sqrt{N}} 
+ \cdots \;\;  ; \;\; \chi(N)=N \Delta_T(N).
\ee
The results for $\chi(N=40)$ and the extrapolated value $\chi$ are shown
in Fig.~\ref{chi_extra}. In a gapful quantum paramagnetic phase, i.e., for
$J_2^{c_1} < J_2 <
J_2^{c_2}$, $\chi$ should vanish. Indeed, we observe a vanishing of $\chi$
around $J_2=0.5J_1$. However, the $J_2$ parameter region where $\chi$ is zero 
is clearly smaller than that where the order parameters $m_0$ and $m_1$
vanish.
At $J_2=0$  we obtain  $\chi=0.0674$ which is in good 
agreement with data obtained by other methods, e.g.,  quantum Monte
Carlo ($\chi= 0.0669$) \cite{qmc}, series expansion ($\chi = 0.0659$)
\cite{zheng},
third-order spin-wave theory ($\chi = 0.06291$) \cite{hamer}, or coupled
cluster method ($\chi = 0.065$) \cite{farnell}.

So far we have considered quantities which can be calculated directly for each finite
lattice and for which the corresponding value for $N \to \infty$
is the leading term in a finite-size extrapolation formula, see
Eqs.~(\ref{m_0}), (\ref{m_1}), (\ref{e}), (\ref{eq_gap}), and (\ref{eq_chi}).
A test of the accuracy the finite-size extrapolation is given by the
comparison with best available data for the unfrustrated limit
($J_2=0$), see above.
In addition to the direct calculation  of certain magnetic quantities the finite
size-extrapolation allows
also an indirect 
determination of the spin-wave velocity $c$ and the spin stiffness $\rho$
which enter the finite size-extrapolation formulas (\ref{m_0}) and
(\ref{e}) as prefactors of the leading finite-size corrections.     
First we can determine the spin-wave velocity $c$ via Eq.~(\ref{e}). Using
this result and also $m_0$ we find then the stiffness $\rho_1 \propto
m_0^2c$ via Eq.~(\ref{m_0}). Since $\rho_1$ is proportional to $m_0^2$, it vanishes at the same point
$J_2^{c_2}$. Alternatively, we can use the hydrodynamic
relation $\chi =\rho/c^2$, see, e.g., Refs.~\cite{chakra} and
\cite{qmc}, to
determine the stiffness $\rho_2=\chi c^2$.  
      We show the spin-wave velocity and both values of the stiffness in Fig.~\ref{c_sw}. 
Although, $\rho_1$ obtained via Eq.~(\ref{m_0}) and $\rho_2$ obtained via
the hydrodynamic relation are quite different the qualitative behavior of
both is similar to that obtained by direct
calculation of $\rho$, e.g., using
Schwinger boson \cite{Trumper97} or coupled cluster approach \cite{darradi08}.
The spin-wave velocity $c$ also decreases with growing $J_2$. However, $c$ 
remains finite at the magnetic-nonmagnetic
transition at $J_2^{c_1}$ as predicted from the non-linear sigma
model \cite{chakra}.
At $J_2=0$ we get $c=1.287$, $\rho_1=0.1246$ and 
$\rho_2=0.1115$. These values are significantly lower than  corresponding
values obtained by quantum Monte Carlo method \cite{qmc}, 
3rd order spin-wave theory \cite{hamer}
or coupled cluster method \cite{ccm,ccm1}. Hence we may argue that the
indirect determination of magnetic quantities via the prefactors of the
leading finite-size corrections is less accurate. 

\section{Summary}
In this paper we have presented detailed information on the GS and
the low-lying excitations for the  spin-$1/2$ $J_1$-$J_2$
HAFM on a finite square lattice of $N=40$ sites. 
These data for the GS  energy, the lowest triplet and singlet
excitations, the spin-spin correlation functions and the magnetic order
parameters may serve as benchmarks for approximate methods.

Including data for finite lattices of $N=20, 32, 36$ sites we have performed
finite-size extrapolations of several magnetic quantities. 
Using the well-known extrapolation formulas for
the GS energy, the order parameters, the spin gap, and the uniform
susceptibility we have determined these quantities for $N \to \infty$.
To estimate the accuracy of the extrapolated quantities we have compared
them with best available results for the unfrustrated limit $J_2=0$, where 
the high-order spin-wave theory and the quantum Monte Carlo method work well.              
At $J_2=0$ the extrapolated GS energy is in excellent agreement with
spin-wave theory and quantum Monte Carlo results. The deviation of the \Neel order
parameter obtained by the finite-size extrapolation from the spin-wave
and quantum Monte Carlo results is only about 1\%.

Besides these quantities which have  been determined directly for each
finite lattice, in addition,  we have used the formulas of finite-size
extrapolation to determine the spin-wave velocity as well as the
spin stiffness which appear in the prefactors of the leading
finite-size corrections of the GS energy and the \Neel order parameter.
This 'indirect way' to determine spin-wave velocity and
stiffness, however, yields only qualitative agreement with known
results.

From the extrapolated magnetic order parameters 
we find the transition point from the magnetically ordered \Neel phase
to the gapful quantum paramagnetic phase 
to $J_2^{c_1} \approx 0.35J_1$ and transition point between the magnetically ordered
collinear phase and the quantum paramagnetic phase   
to $J_2^{c_2} \approx 0.66J_1$.\\

{\it \bf  Acknowledgment:}
This work was supported by the DFG (Ri615/16-1).


\end{document}